\documentclass[runningheads]{llncs}

\usepackage[utf8]{inputenc} 
\usepackage[T1]{fontenc}

\usepackage{amssymb}
\usepackage{pifont}
\usepackage{tabularx}
\usepackage{multirow}

\usepackage{graphicx} 

\usepackage{booktabs}

\usepackage{comment}
\usepackage{todonotes}

\title{\LARGE \bf Look: AI at Work! --\\ Analysing Key Aspects of\\ AI-support at the Work Place}

\author{%
  Stefan Schiffer\inst{1,2}%
  \orcidID{0000-0003-1343-7140} %
  \and 
  ~\\
  Anna Milena Rothermel\inst{2}%
  \orcidID{0009-0005-2377-9277} %
  \and 
  ~\\
  Alexander Ferrein\inst{1}%
  \orcidID{0000-0002-0643-5422} %
  \and 
  ~\\
  Astrid Rosenthal-von~der~P{\"u}tten\inst{2}
  \orcidID{0000-0002-2497-143X}%
}

\authorrunning{S. Schiffer et al.}

\institute{%
  Mobile Autonomous Systems and Cognitive Robotics (MASCOR) Institute,\\
  FH Aachen University of Applied Sciences, Aachen, Germany
  \email{\{s.schiffer,ferrein\}@fh-aachen.de}%
  \and
  Chair Individual and Technology (iTec), RWTH Aachen University, Aachen Germany
  \email{\{stefan.schiffer, milena.rothermel, arvdp\}@itec.rwth-aachen.de}%
}

\begin{document}

\maketitle

\thispagestyle{empty}
\pagestyle{empty}

\begin{abstract}
  In this paper we present an analysis of technological and psychological factors
  of applying artificial intelligence (AI) at the work place. %
  We do so for a number of twelve application cases in the context of a project where AI is integrated at work places and in work systems of the future.
  From a technological point of view we mainly look at the areas of AI that the applications are concerned with.
  This allows to formulate recommendations in terms of what to look at in developing an AI application and
  what to pay attention to with regards to building AI literacy with different stakeholders using the system.
  This includes the importance of high-quality data for training learning-based systems
  as well as the integration of human expertise, especially with knowledge-based systems.
  In terms of the psychological factors we derive research questions to investigate
  in the development of AI supported work systems and to consider in future work,  
  mainly concerned with topics such as acceptance, openness, and trust in an AI system.
  
  %
\end{abstract}

\section{Introduction}

In this paper, we report lessons learnt from developing applications of Artificial Intelligence (AI) at the work place.
Our work takes place in the context of the project WIRKsam \cite{Jeske:etAl_LuE2023WIRKsam_Projektvorstellung}.
%
In a collaboration with partners from ergonomics, work science,
computer science and psychology the project tries to identify and use the
potential of AI support at the work place. %
In a participatory 
fashion the partners look at how to apply AI to improve work places
and work systems in order to increase the satisfaction of workers and
other stakeholders and to enhance the value of such work places. %

This interdisciplinary research features different perspectives,
including ergonomics, human factors, and technological as well as
psychological aspects. %
In this paper, we only report on the technological and the
psychological factors. %
Drawing from the experiences in developing AI application in twelve use
cases we present lessons learnt and share insights about what needs to
be taken into account and what are interesting research questions when
applying AI at the work place.
%
In terms of the technological factors we look at the sub-fields of AI that the particular applications call for.
Since every technique has its own set of requirements and implications, we can use this to
derive a prioritization for building specific forms of AI literacy with different stakeholders correspondingly.
%
Human factors include things like usability, (perceived) ease-of-use, acceptance, intention to use. %
What is more, we are interested in (social) psychological factors, such as perceived fairness,
self-perception, changes in social dynamics, for instance with regards to hierarchies, or diffusion of responsibility.

The rest of the paper is organized as follows. %
We will first lay out the background of the project before we start
our analysis of the technological aspects. %
Then we turn to aspects of (social) psychology. %
Finally, we discuss possible implications for successfully
deploying AI at the work place and directions for future research.

\section{Background} 


Industry in Germany is facing a number of challenges to stay successful. %
Apart from a general pressure due to high energy costs,
high wages levels, and the international competition especially in
production, companies in general struggle with a shortage of skilled
workers, and small and medium enterprises do so in particular.%
\footnote{\url{https://www.bitkom.org/Presse/Presseinformation/Kuenstliche-Intelligenz-2022}}
One reason, for example, is that the oversupply of skilled jobs leads
to workers denying to work night shifts and such. %
Sometimes companies even need to cut down production for that
reason \cite{Burstedde-Kolev-Schaefer_IW2024-27_Kosten-FKM}.

A possible way to counteract this is to reach high(er) levels of automation in order to keep output high.%
\footnote{\url{https://www.bcg.com/publications/2023/scaling-ai-pays-off}} %
At the same time, in the public (supposed) breakthroughs in AI
technology pressure companies to make use of these new technologies,
while even skilled workers fear for their jobs. %
That is why, when introducing AI technology, one needs to carefully
develop solutions together with the workers and one needs to convince
them that their job is not (necessarily) at risk, but that AI can
provide a useful tool to make jobs easier or more enjoyable for them.

\subsection{Motivation}

The introduction of useful AI tools and support needs to be thought of from a technical and a socio-technical side.
On the technical side, AI technology needs to be developed with workers in the loop and still in charge.
Further, workers need to be instructed and taught AI principles to better understand the ``AI'' system(s) they are working with.
They need a basic form of AI literacy in order to be able to demystify the systems that are supposed to support them.
On a socio-technical side one needs to investigate and work out the benefits and challenges in human-technology interaction.
Among others, this includes acceptance of the system, openness to use/integrate it,
trust in its work/outcome, or perceived fairness of automated decisions.
In order to provide lessons learnt we draw from experiences in applying AI in a number of use cases in a large scale project.

\subsection{The WIRKsam Project}

The WIRKsam project 
looks at the application of AI at the work place in twelve use cases.
A very brief description of the use cases is given in Table~\ref{tab:usecases}.
For a more detailed account of these applications we refer the interested reader to \cite{Leistung-und-Entgelt_2024_WIRKsam}.

\begin{table}[tbp]
  \centering
  \caption{Overview of the different use cases in WIRKsam}
  \begin{tabularx}{\linewidth}{>{\bfseries}lX}
    \toprule
    \normalfont{Use Case}  & Objective \\ 
    \midrule
    1.1 & multi-criteria optimization for process control \& qualification of products\\
    1.2 & AI expert system for acurate production of 3D textiles\\
    1.3 & AI-based support of skill development in metal forming\\
    1.4 & quality control and error type classification\\
    2.1 & variability management in fabric production\\
    2.2 & demand forecast for reduced stock-keeping\\
    2.3 & AI-based manufacturing control in metal working\\
    2.4 & process planning in machine assembly\\
    3.1 & optimized quality control for automotive filters\\
    3.2 & parameter estimation for analysis of wear of press felts\\
    3.3 & robot-assisted fabrication of fiber-reinforced composites\\
    3.4 & joint categorization and cutting scrap planning\\
    \bottomrule
  \end{tabularx}
  \label{tab:usecases}
\end{table}

The WIRKsam project follows a procedure model described in \cite{harlacher2023approach}.
The model tries to join work science-related consideration with technical considerations,
loosely coupling the two sides with one another. %
The technical parts mainly follow the CRISP-DM model \cite{Wirth:Hipp_PAKDDM2000_CRISP-DM}. %
In WIRKsam, results are achieved iteratively and presented transparently.
This is done in a living lab and by means of so-called demonstrators, which are prototypes of the AI systems,
that are planned for explicitely \cite{Hansen-Ampa-etAl_IEA2024_ConceptualizationDemonstrators}.
Especially in the research of social psychological factors the use of such demonstrators is an important opportunity
to gain insights into how humans interact with technology and how affected they feel by the new technology.

\subsection{Example Cases}
\label{subsec:example-cases}

Instead of describing or analysing every single use case mentioned in
Table~\ref{tab:usecases}, we take three prototypical examples 
and discuss it in more detail.
These three example use cases show 
interesting characteristics that can also be found in other use cases as well.
\begin{description}
\item[I.] multi-criteria optimization for process control \& qualification of products\\
  In the production of an implantable medical textile a large number
  of parameters influence the quality of the yarn.  Implicit knowledge
  of the person operating the machine should be objectified and
  modelled in a comprehensible way. This data can then be used to
  help optimize machine parameter tuning based on the knowledge of
  experienced workers.
\item[II.] AI-based manufacturing control in metal working\\
  As part of a weekly production planning meeting adjustments are being made
  to which products can be produced given the current situation of
  materials in stock, workers with appropriate skills available and
  machines operational. The decision about which adjustments to make
  should be supported by an AI system that projects the consequences
  of changes to an existing production plan and helps to accommodate
  for new situations appropriately.
\item[III.] optimized quality control for automotive filters\\
  In the quality control of automotive metal filter meshes currently
  every product is manually inspected for deficiencies. %
  In order to ease the work for humans in quality control an AI system
  should support inspection by means of computer vision. %
  A description of the technical solution is given in
  \cite{Arndt:etAl_PETRA2023_AnomalyDetection}.
\end{description}


\subsection{Technological Factors}

The term \emph{Artificial Intelligence} (AI) was coined by John McCarthy
and others in their research proposal from 1955~\cite{McCarthy:etAl_1955_Proposal-for-Dartmouth}
(reprinted in \cite{McCarthy_Minsky_Rochester_Shannon_AI-Mag2006}). %
Since there is no single definition of AI and different interpretations of the term exist, 
%
the European Union formed a High-level Expert Group
to create a definition of AI \cite{AI-HLEG_EU2019_DefinitionAI}. %
to achieve a common understanding as a basis for further cooperation.
In the following we understand AI as the set of techniques and methods
that allow a computer system (or other technical artifact) to behave intelligently.
This is largely based on a definition from~\cite{Nilsson_1998_Artificial-Intelligence}. %

There is a large body of different methods and techniques
to approach this, spanning many different sub areas of AI.
An extensive description is provided in~\cite{Russell:Norvig_2020_AIMA}.
%
In this regard,
Shapiro reflects on the concept of AI-completeness \cite{shapiro2003artificial}.
Other overviews of (sub-)areas of AI are mentioned in 
\cite{Kersting:Peters:Rothkopf_2019_WasIstEineProfessurFuerKI} and
\cite{Humm_2020_AppliedArtificialIntelligence}.
Techniques from different sub areas show quite requirements in terms of
what one needs to know to be able to use a particular methods (e.g. domain specific knowledge),
what has to be provided as an input (e.g. training data) and
what needs to be made transparent to interpret the output of an algorithm.
That is why we will look at which methods appear in our use cases frequently
and which implications that might have for the human-machine interaction.

\subsection{Psychological Factors}

Integrating AI in the work place potentially causes the humans’ work environment and tasks to change.
The psychological factors that arise need to be examined in order to successfully apply AI.
Therefore, it is essential to investigate the humans’ acceptance, openness, and trust in both, the AI in general and its work.
Regarding AI as a decision making system for example, it is important to examine the extent to which humans trust the AI’s suggestions.
This ranges from blindly trusting the AI without questioning the decision to not trusting the AI at all and in turn not profiting from its capabilities.
The degree of trust that is appropriate depends on the respective AI and its competences.
Openness, in this context, reflects the humans’ openness to use AI.
It might influence the intention to use AI as well as it could be influenced by other factors such as knowledge about AI \cite{lautenschlaeger2020} 
or attitudes towards AI \cite{doi:10.1080/00223980.2021.2012109}.
Moreover, the degree to which humans would use AI is positively related to the acceptance of AI \cite{Sindermann2021}, 
attitudes towards AI \cite{KELLY2023101925} 
and trust in AI \cite{KELLY2023101925}. 
Acceptance of AI could be measured using the Attitude Towards Artificial Intelligence Scale, which consists of two subscales assessing acceptance and fear of AI, respectively  \cite{Sindermann2021}. 
Trust in AI on the other hand could be measured using the Trust Scale for the AI context \cite{10.3389/fcomp.2023.1096257,scharowski2024trustdistrusttrustmeasures}. 

When it comes to the AI making personnel or task assignment related decisions it might also be interesting to look at its perceived fairness \cite{Zhou2022}.
In this regard, not only the fairness of the system itself but also in comparison to a human colleague should be taken into account.
In various applications of AI, the preference for either an AI or humans fulfilling the tasks could be examined.
Furthermore, regarding the communication of AI with humans, the extent to which an AI offers explanations about a decision and its formation should be analyzed considering different areas of application of AI. The amount of explanation provided by the AI might for instance affect the perceived fairness of and trust in a system \cite{Zhou2022}.

Due to the AI taking over tasks that were formally completed by humans, their self-perceptions within the work context might change, as well.
Therefore, their perceived competence, autonomy, control, and self-efficacy ought to be explored, as well as changes in professional identity or roles at work.
With the AI handling potentially undesirable or even dangerous tasks, another factor that should be addressed is the perception of the work itself.
This might include alterations in one’s wellbeing, along with the perceived meaningfulness and monotony of tasks.
Thus, not only the feelings and attitudes towards an AI should be considered but also the changes in self- and task-perception.
These might in turn affect attitudes toward AI and therefore be crucial for the successful integration of AI at the workplace.

\section{Technological Factors}



In order to derive recommendations from a technological point of view,
we start by categorizing our applications in terms of the type(s) of AI
method/technique that are needed or feasible to solve the problem at hand. %
We do so by following a classification scheme given in
\cite{Schmid:etAl:KI2021:AIMCsquare}. %
It is similar to other classifications like mentioned in
\cite{Kersting:Peters:Rothkopf_2019_WasIstEineProfessurFuerKI} or
\cite{Humm_2020_AppliedArtificialIntelligence} and the areas of AI
covered in \cite{Russell:Norvig_2020_AIMA}.

The main categories from \cite{Schmid:etAl:KI2021:AIMCsquare} with their first sub-level are as follows.
\begin{description}
\item[Problem solving, Optimization, Planning, and Decision Making]~\\ %
  Problem Solving, Optimization, Planning and Plan Recognition, Decision Making
\item[Knowledge Representation and Reasoning]~\\ %
  Knowledge Representation, Reasoning, Uncertain Knowledge,
  Probabilistic Reasoning, Non-probabilistic Reasoning,
  Other Approaches for Uncertain Reasoning
\item[Machine Learning]~\\ %
  Supervised Learning, Unsupervised Learning,
  Semi-supervised Learning, Reinforcement Learning
\item[Hybrid Learning]~\\ %
  Hybrid Neural Systems, Learning with Knowledge, Conversational Learning
\end{description}

Instead of classifying every single use case we rather collect those areas where comparably many of the use cases potentially have their solution.
Sub-areas that appear in at least three of the use cases seem to be relevant since that makes for already 25\% of the applications then.
Following this analysis we see that seven areas are touched by the most cases.
It is all of the areas in the first block, that is Problem Solving, Optimization, Planning and Plan Recognition, and Decision Making.
Also, many cases are concerned with Machine Learning, both in form of supervised but also unsupervised machine learning.
Finally, probabilistic reasoning is an important topic with some cases.
In the selection of the example cases in Section~\ref{subsec:example-cases} we tried to cover as many of the seven areas.


We see that many applications feature some form of decision support (e.g. expert systems, recommender systems).
That is why in developing AI applications we need to evaluate the availability of data for the human to be able to accept (or reject) the AI system's recommendation/suggestion.
  This concerns our example cases I and II. %
  For the example case I, such an expert system can be used to transfer expert knowledge from one generation of workes to the next.
  In example case II the decision support could try to objectify the decisions made by humans and to make them more tranparent to those affected by it.
  
  Because many applications use (supervised) machine learning techniques (example cases I and III), the value of proper input data for training has to be stressed.
  Companies have to be sensitized for the importance of providing a sufficient amount of (high-quality) data since the quality of the resulting models depends on it.
  In our experience this is often underestimated, both in terms of the effort needed and the possible benefits.
  For those cases that use unsupervised learning techniques it is important to involve human experts in interpreting the models.
  Combining the learnt models with human insights allows for lifting their full potential.

  A number of cases have to do with optimization (example case I).
  An advantage of t an AI solution here ist that it can compare a larger number of alternatives than humans could do.
  Similar reasons hold for (process) planning (example case II). An
  algorithm can evaluate many more different solutions to a problem
  than a human could ever do.


\section{Psychological Factors}

Within the previously described use cases, psychological factors arise that should be investigated.
As for example use case I, the AI is supposed to handle the settings of a complex machine and help specialists with the control of fabric production.
This means that the employee's responsibilities are about to change.
That is why in this use case the self-perception of the employees particularly comes into focus.
On the one hand, changes in the perceived competence and self-efficacy should be taken into consideration.
Previously, a specialist was responsible for setting the machine and controlling the output and therefore had specific knowledge and expertise that might have contributed to feeling competent and self-efficient.
If an AI takes over these tasks and responsibilities it might simultaneously take away the employee's feeling of competence.
On the other hand, 
providing the AI with expert knowledge that is unique to the specialist might be another source for feeling competent and valuable.
Additionally, this deviating distribution of tasks might result in a shift in professional identity and roles.
%
Further psychological factors that could be assessed are the perceived autonomy and control.
Trusting an AI with critical tasks might reduce the feeling of control and autonomy.
However, it could also increase the feeling of having everything under control because of the AI's competence and reassurance.
All of these factors are fundamental to investigate for developing an AI dependent on the knowledge of specialists and for eventually integrating the AI into the workplace with keeping the employees wellbeing and job satisfaction in mind.

In example case II %
the AI is meant to assist in shift scheduling and task coordination.
As in this use case the AI is designed to decide about the coordination of human employees, the perceived fairness of the system is of particular importance.
Research should investigate how employees evaluate the fairness of AI decisions and whether they prefer a human or an AI making the decisions.
Another factor is the acceptance of the decisions and whether employees are more prone to accept human or AI decisions.
On the one hand the AI could be seen as more objective and logical and therefore fairer than a human.
On the other hand a human might be perceived as more understanding and capable of rating the importance of personal commitments.
The respective research could contribute to designing an AI that not only makes the task and shift assignment more efficient but that also makes the employees perceive it as fairer and more objective, which in turn might increase acceptance and satisfaction.

Regarding example case III, %
the AI is intended to take over physically and mentally demanding quality control tasks.
Psychological factors that should be considered are acceptance and trust in the AI.
Employees that were previously responsible for examining the products are asked to rely on the AI's competence to make the right decision.
Trust and acceptance of the AI need to be ensured for the employees to benefit from the AI's competences and from improved working conditions.
After a successful integration of AI, the employees wellbeing, job satisfaction, and workload could be evaluated.

\section{Conclusion}

In this paper, we reported on different use cases deploying AI technology in real-world SME applications.
Technology-wise, the different use cases are from different AI areas including problems from the sub-area of problem solving, optimization, planning, and decision making as well from unsupervised machine learning.
Finally, probabilistic reasoning plays an important role in our real-world use cases.
When we launched a project, defining the objectives as well as the available date sources following an extended CRISP-DM model was of great help.
It helped not only for the finding the right AI technology, but also to focus the SME on the objectives in their particular use case.
Indeed, it turned out that the availability of high-quality data for machine learning approaches often was an issue.
Another important driver was the basic understanding of the deployed methods on the SME side which helped to make quick progress.
%
On the psychological side, humans’ acceptance, openness, and trust of the AI deployed are of importance.
For recommender systems, it is important to track to which extend the human workers trust the suggestions made by the AI.
In addition, openness is of utter importance; if the workers are not open to give the AI system a chance to prove its usefulness,
as with many other technological systems that have been introduced top-down.
It is therefore important from the direct beginning to integrate the workers in the development process of the AI system to build acceptance and trust into the system.
This also goes hand in hand with understanding how the AI methods works for the workers,
to also better understand the limitations of the system.
Within the WIRKsam project we have an interesting playing field to investigate the mentioned interesting questions
and to better learn, how acceptable AI systems need to be designed.

Concluding, our core findings are as follows.
From a technological perspective, the importance of proper high-quality data for training
and the integration of human expertise is central to building successful applications.
Also, building at least basic AI literacy is crucial for enabling AI support at the work place.
The central psychological factors are acceptance, openness, and trust in an AI system, that appear in various workplace settings.
Thus investigating the extent to which humans accept a newly introduced AI and evaluate its performance compared to a human’s performance is of great importance.
Additionally, an AI system taking over tasks might affect the humans’ professional identity and self-perception, which could in turn be crucial factors for acceptance and openness to AI.
In general terms, a participatory and iterative design and development of AI support at the work place with early and continuous involvement of all stakeholders appears to be a valuable proceeding to integrate the aforementioned aspects.

\subsection*{Acknowledgments}

We acknowledge the support by the Federal Ministry of Education and Research (BMBF) under grant numbers 02L19C601 and 02L19C602. %
Further, we thank the anonymous reviewers for their valuable comments.

\clearpage
\bibliographystyle{splncs04}
\bibliography{biblio}

\end{document}